\documentclass[twocolumn,showpacs,aps,prl,a4paper]{revtex4}

\usepackage{amsmath,graphicx}

\begin{document}

\title{Continuous-Variable Quantum Teleportation with a Conventional Laser
\footnote{This is a revised version of quant-ph/0301045. The author has not replaced that original version with this revised one so that readers can easily verify priority of ideas first presented in the original version in every arXiv mirror site.}}

\author{Mikio Fujii}
\email[Email address: ]{mikio.fujii@toshiba.co.jp}

\affiliation{Systems Integration Technology Center, Toshiba Corporation, 3-22 Katamachi, Fuchu, Tokyo, 183-8512, Japan}

\begin{abstract}
We give a description of balanced homodyne detection (BHD) using a conventional laser as a local oscillator (LO), where the laser field outside the cavity is a mixed state whose phase is completely unknown. Our description is based on the standard interpretation of the quantum theory for measurement, and accords with the experimental result in the squeezed state generation scheme. We apply our description of BHD to continuous-variable quantum teleportation (CVQT) with a conventional laser to analyze the CVQT experiment [A. Furusawa {\it et al.}, Science {\bf 282}, 706 (1998)], whose validity has been questioned on the ground of intrinsic phase indeterminacy of the laser field [T.\ Rudolph and B.C.\ Sanders, Phys.\ Rev.\ Lett.\ {\bf 87}, 077903 (2001)]. We show that CVQT with a laser is valid only if the unknown phase of the laser field is shared among sender's LOs, the EPR state, and receiver's LO. The CVQT experiment is considered valid with the aid of an optical path other than the EPR channel and a classical channel, directly linking between a sender and a receiver. We also propose a method to probabilistically generate a strongly phase-correlated quantum state via continuous measurement of independent lasers, which is applicable to realizing CVQT without the additional optical path.
\end{abstract}

\pacs{03.67.Hk, 03.65.Ud, 42.50.Dv}

\maketitle

Quantum teleportation is a method to move quantum states from a sender ``Alice'' to a receiver ``Bob'' by the aid of entanglement. The original protocol \cite{Original QT} has later been extended to continuous-variable quantum teleportation (CVQT) using a two-mode squeezed state and balanced homodyne detection (BHD) \cite{CVQT}.
While experimental demonstration of CVQT was reported as the first achievement of unconditional quantum teleportation \cite{CVQT Experiment}, there has been controversy over its validity on the ground of intrinsic phase indeterminacy of the laser field \cite{RS, vEF}.
The laser field is often assumed to be a coherent state having a fixed phase, but the steady-state solution of the master equation in the quantum theory of the laser shows that the phase of the laser field inside the cavity is completely unknown in operation well above threshold \cite{RS, vEF, Walls and Milburn, Moelmer}.

In this Letter, we first discuss a description of the laser field outside the cavity in line with the previous discussions \cite{RS, vEF}. We then analyze BHD using a laser as a local oscillator (LO), and give a physically appropriate description of BHD based on the standard interpretation of the quantum theory for measurement. We apply our description of BHD to CVQT with a laser to analyze the CVQT experiment \cite{CVQT Experiment}. Finally, we propose a method to probabilistically produce a strongly phase-correlated quantum state via continuous measurement of independent lasers, which is applicable to realizing CVQT without an optical path between Alice and Bob for sharing the same laser field.

In Ref.\ \cite{vEF}, van Enk and Fuchs claim that the standard description of the laser field used in Ref.\ \cite{RS} is surprisingly insufficient to understand CVQT with a laser. By employing the input-output theory \cite{Walls and Milburn}, they find that the laser field outside the cavity is a continuous-mode mixed state which is in form identical to the steady-state field inside the cavity. They then go on to express the field state in terms of noncontinuous operators given by them, and discuss its coherency.

The definition of noncontinuous operators is based on the narrow bandwidth approximation (NBA) $B \ll \omega_0$, where $\omega_0$ is the central frequency of the bandwidth $B$. Under NBA, the annihilation operator part of the continuous-mode electric field is approximated as $\hat{E}^{+}(z,t) \sim i \mathcal{E}_0 \sum_n \phi_n(t-z/c) \hat{c}_n$, where $\mathcal{E}_0 \equiv [\hbar\omega_0/(2\epsilon_0 c A)]^{1/2}$, $\{\phi_n(t)\}$ are basis functions giving a profile of the fundamental modes, and $\hat{c}_n$ is a noncontinuous annihilation operator \cite{BLP}.

In Ref.\ \cite{vEF}, a subset of basis functions $\{\psi_n(t)\}$ is specifically chosen as $\psi_n(t) \equiv (T)^{-1/2} \exp(-i\omega_0 t)\Pi(t/T-n)$ where $\Pi(t)$ is the rectangle function, and importance of the field expression in terms of noncontinuous operators defined by $\{\psi_n(t)\}$ is emphasized in view of the quantum de Finetti theorem.
But this given expression is nothing more than one possible approximation of the field state outside the cavity by NBA, where the exact state obtained by the input-output theory is actually identical to the standard description of the laser field used in Ref.\ \cite{RS}.

We will show that, contrary to the claims of Ref.\ \cite{vEF}, the standard description of the laser field used in Ref.\ \cite{RS} is sufficient to understand CVQT with a laser.
Throughout this Letter, we use the standard description of the laser field in Ref.\ \cite{RS} for the continuous mode outside the cavity.
We explain coherency of the laser field not by interpreting a new approximate expression of the laser field as Ref.\ \cite{vEF}, but by appropriately formulating measurement process for the laser field.

Now, we will give a description of BHD with a conventional laser.
As long as the photon number operator well represents an observable for an efficient photodetector lacking single photon resolution \cite{Carmichael, Gardiner and Zoller}, we may regard $\hat{a}_l^\dagger \hat{a}_s + \hat{a}_l \hat{a}_s^\dagger$ as an observable for BHD, where $\hat{a}_l$ and $\hat{a}_s$ are annihilation operators for the LO field and the signal field, respectively \cite{Braunstein}.
If the signal field $|\psi \rangle_s$ satisfies $r \sqrt{\langle \psi | \mbox{\small $\hat{X}_s(\theta)^2$} | \psi \rangle_s} \gg \sqrt{\langle \psi|\mbox{\small $\hat{a}_s^\dagger\hat{a}_s$}|\psi\rangle_s}$, which holds when the intensity of the LO field is extremely larger than that of the signal field, this observable satisfies
\begin{eqnarray}
(\hat{a}_l^\dagger \hat{a}_s + \hat{a}_l \hat{a}_s^\dagger) |r e^{i\theta} \rangle_l |\psi \rangle_s&\sim& r \hat{X}_s(\theta) |r e^{i\theta} \rangle_l |\psi \rangle_s, \label{eigenvalue equation}
\end{eqnarray}
 where $\hat{X}(\theta) \equiv \hat{a} e^{-i\theta} + \hat{a}^\dagger e^{i\theta}$ and $|re^{i\theta}\rangle$ is the coherent state in polar coordinates \cite{Walls and Milburn}.
According to the standard interpretation of the quantum theory \cite{Sakurai}, Eq.\ (\ref{eigenvalue equation}) implies if we obtain the measurement outcome $rx$ in one trial of BHD with the prior knowledge of $r$, $|\psi\rangle_s$ instantaneously reduces to $|x, \theta \rangle_s$ satisfying $\hat{X}_s(\theta)|x, \theta \rangle_s = x |x, \theta \rangle_s$.

Since $r$ of the laser field is measurable beforehand, we may define the measurement operator \cite{Nielsen and Chuang} for BHD as
\begin{eqnarray}
\hat{M}(x, r, \theta) \equiv \pi^{-\frac{1}{2}} \; |r e^{i\theta}\rangle_l|x, \theta \rangle_s \langle x, \theta |_s \langle r e^{i\theta}|_l , \label{BHD: measurement operator}
\end{eqnarray}
where $|x, \theta \rangle$ is the quadrature eigenstate written as
\begin{eqnarray}
|x, \theta \rangle
&=& (2\pi)^{-\frac{1}{4}} e^{-\frac{x^2}{4}}
\exp\left( x e^{i\theta}\hat{a}^{\dagger}-\mbox{$\frac{1}{2}$}e^{i2\theta}\hat{a}^{\dagger2} \right)|0\rangle . \quad   \label{quadrature eigenstate}
\end{eqnarray}
$|x, \theta \rangle$ satisfies the orthonormalization condition $\langle x_1, \theta | x_2, \theta \rangle = \delta(x_1 - x_2)$ and the completeness relation $ \int_{-\infty}^{+\infty} dx \ |x, \theta \rangle \langle x, \theta | = 1 $ on $x$. Since the coherent state also satisfies the completeness relation \cite{Walls and Milburn}, Eq.\ (\ref{BHD: measurement operator}) satisfies the completeness relation
$
\int_{-\infty}^{+\infty} dx \int_0^{\infty}r dr \int_0^{2\pi} \!\! d\theta \ \hat{M}^{\dagger}(x, r, \theta)\hat{M}(x, r, \theta) = 1
$.

Considering that we cannot distinguish between $|x, \theta_1 \rangle$ and $|x, \theta_2 \rangle$ ($\theta_1 \neq \theta_2$) by measurement results due to intrinsic phase indeterminacy of the laser field \cite{RS, Moelmer}, the probability of obtaining the measurement outcome $x=\bar{x}$ with the prior knowledge of $r$ is
\begin{eqnarray}
P(\bar{x}) = \int_0^{\infty} \!\!\!\! r dr \int_{0}^{2\pi} \!\!\!\!\! d\theta
\  \mathrm{Tr} \left\{ \hat{M}(\bar{x}, r,  \theta) \hat{\rho}_o  \hat{M}^\dagger (\bar{x}, r , \theta)\right\}, \label{OBPM: probability}
\end{eqnarray}
and the density operator after the measurement is
\begin{eqnarray}
\hat{\rho}
= P(\bar{x})^{-1} \!\!\! \int_0^{\infty} \!\!\!\! r dr \int_{0}^{2\pi} \!\!\!\!\! d\theta  \; \hat{M}(\bar{x}, r,  \theta) \hat{\rho}_o \hat{M}^\dagger (\bar{x}, r,  \theta), \label{OBPM: dentity operator}
\end{eqnarray}
where $\hat{\rho}_o$ is the density operator before the measurement.

We will denote the procedure described above the observable-based projection method (OBPM) in the rest of this Letter.
Note that above discussion is not based on the assumption that the laser field is the coherent state (``partition ensemble fallacy'' \cite{RS, KB}).
It is the property of the observable for BHD that approximately projects the strong laser field of the LO mode onto the coherent state after the measurement. On the contrary, the number states in the LO mode cannot be eigenstates of the observable for BHD, because $|n\rangle \neq |n-1\rangle$ even in the limit $n \to +\infty$ due to their rigid orthogonality.

As an example of BHD, we will calculate $P(\bar{x})$ in the squeezed light generation scheme \cite{Squeezed State Experiment} by OBPM. In the scheme, the same laser source is used for supplying the LO field, and pumping the nonlinear medium to generate the squeezed state.
The density operator of the system before the measurement is
\begin{eqnarray}
\hat{\rho}_o \!=\!\! \int_{0}^{2\pi} \!\!\! \frac{d\phi}{2\pi} \ |r_{o} e^{i(\phi+\varphi)}\rangle_l|0,se^{i2\phi}\rangle_s
\langle 0, se^{i2\phi}|_s \langle r_{o} e^{i(\phi+\varphi)}|_l, \quad \label{SQ: initial density operator}
\end{eqnarray}
where $\phi$ is the unknown phase of the pump field, $\varphi$ is the phase delay by a controllable phase shifter, and $|0, \varepsilon \rangle \equiv \hat{S}(\varepsilon)|0 \rangle$ is the squeezed vacuum state \cite{Walls and Milburn}. The unknown phase of the squeezed state is $2\phi$ instead of $\phi$, because frequency of the pump field is doubled by second harmonic generation before the field enters an optical parametric oscillator.
By using Eqs.\ (\ref{OBPM: probability}), (\ref{SQ: initial density operator}), orthogonality approximation of the coherent state $|\langle r e^{i\theta} | r_o e^{i\theta_o} \rangle|^2 \sim (\pi/r_o)\delta(r-r_o)\delta(\theta-\theta_o)$ in the limit $r_o \to +\infty$ derived from $\lim_{\epsilon \to 0+} \exp[-t^2/(4\epsilon)]/(2\sqrt{\pi\epsilon}) = \delta(t)$, and the relation
\vspace{-2.5mm}
\begin{eqnarray}
\lefteqn{\langle x, \theta | 0, s e^{i2(\theta-\varphi)} \rangle
= \sum_{n=0}^{\infty} \langle x, \theta | n \rangle \langle n | 0, s e^{i2(\theta-\varphi)} \rangle} \nonumber \\
&=& \sum_{n=0}^{\infty} (2\pi)^{-\frac{1}{4}}(2^n n!)^{-\frac{1}{2}} H_n\left(\frac{x}{\sqrt{2}}\right) e^{-\frac{x^2}{4} -in\theta} \hspace{10mm} \nonumber \\
& & \hspace{3mm} \times  [2^n n! \cosh(s)]^{-\frac{1}{2}} [e^{i2(\theta-\varphi)}\tanh(s)]^{\frac{n}{2}}H_n(0) \nonumber \\
&=& \langle x, \varphi | 0,s \rangle \nonumber
\end{eqnarray}
where $H_n(x)$ are Hermite polynomials, we find $P(\bar{x})=|\langle \bar{x}, \varphi | 0, s \rangle|^2$, which agrees with the experimental result of Ref.\ \cite{Squeezed State Experiment}.

Next, we will apply OBPM to CVQT with a laser \cite{CVQT, RS}. In the measurement step by Alice, the probability of obtaining $\bar{x}_1$ in BHD1 and $\bar{x}_2$ in BHD2 is
$P(\bar{x}_1, \bar{x}_2) \equiv \int_0^\infty \! r_1 dr_1 \! \int_0^{2\pi} \! d\theta_1 \! \int_0^\infty \! r_2 dr_2 \! \int_0^{2\pi} \! d\theta_2 \> \mathrm{Tr} \{ \hat{M}_2 \hat{M}_1 \hat{\rho}_\mathrm{I} \hat{M}_1^\dagger \hat{M}_2^\dagger \}$
 and the density operator after the measurement is
$
\hat{\rho}_\mathrm{II} \equiv P^{-1}(\bar{x}_1, \bar{x}_2) \! \int_0^\infty \!\!\! r_1 dr_1 \! \int_0^{2\pi} \!\!\! d\theta_1 \! \int_0^\infty \!\!\! r_2 dr_2 \! \int_0^{2\pi}  \!\!\! d\theta_2 \; \hat{M}_2 \hat{M}_1 \hat{\rho}_\mathrm{I} \hat{M}_1^\dagger \hat{M}_2^\dagger
$,
where $\hat{M}_j \equiv \pi^{-1/2} |r_j e^{i\theta_j}\rangle_{lj}|\bar{x}_j, \theta_j \rangle_{sj}\langle \bar{x}_j, \theta_j|_{sj} \langle r_j e^{i\theta_j}|_{lj} \; (j=1, 2)$. $\hat{\rho}_\mathrm{I}$ is the density operator of the total system before the measurement written as
\vspace{-2mm}
\begin{eqnarray}
\hat{\rho}_\mathrm{I} &=& \int_0^{2\pi} \!\!\! \frac{d\phi}{2\pi} \ |r_o e^{i\phi} \rangle_{l1} |r_o e^{i(\phi+\frac{\pi}{2})} \rangle_{l2} |\eta e^{i2\phi} \rangle_{1,2} \nonumber\\
& & \hspace{15mm} \otimes |r_o e^{i\phi} \rangle_{l3} \, \hat{\rho}_{in} \, \langle r_o e^{i\phi} |_{l3}  \nonumber\\
& & \hspace{10mm} \otimes \langle \eta  e^{i2\phi} |_{1,2}  \langle r_o e^{i(\phi+\frac{\pi}{2})} |_{l2}
\langle r_o e^{i\phi} |_{l1} , \label{CVQT: initial density operator}
\end{eqnarray}
where the modes $l1, l2$ are for LOs of BHD1,2 in Alice, $l3$ for LO in Bob, $\phi$ is the unknown phase of the pump field, $\hat{\rho}_{in}$ is an arbitrary density operator supplied by a third party ``Victor'' to Alice, and $|\eta e^{i2\phi} \rangle_{1,2} \equiv \sqrt{1-\eta^2} \exp ( \eta e^{i2\phi} \hat{a}_1^\dagger \hat{a}_2^\dagger ) |0 \rangle_1|0 \rangle_2$ is a two-mode squeezed state \cite{Walls and Milburn} as the EPR state. Again, the unknown phase in the modes $1, 2$ is $2\phi$ instead of $\phi$. (See Fig.\ 1 in Ref.\ \cite{RS}.)

By using Eq.\ (\ref{quadrature eigenstate}) and $\hat{a}_{s1} = (\hat{a}_{in}-\hat{a}_1)/\sqrt{2}$, $\hat{a}_{s2} = (\hat{a}_{in} + \hat{a}_1)/\sqrt{2}$ where the modes $s1,s2$ are for the signal field of BHD1,2, the quadrature eigenstates of the modes $s1, s2$ are written in the modes $in, 1$ as
$
\left|\bar{x}_1, \phi \right\rangle_{s1} | \bar{x}_2, \phi+\mbox{$\frac{\pi}{2}$} \rangle_{s2}
= [\exp(\mbox{\scriptsize $-\frac{|\gamma|^2}{2}$})/\sqrt{2\pi}] \exp [(\gamma\hat{a}_{in}^\dagger-\gamma^*\hat{a}_1^\dagger)e^{i\phi} + \hat{a}_{in}^\dagger \hat{a}_1^\dagger e^{i2\phi}] |0\rangle_{in} |0\rangle_1,
$
where $\gamma \equiv (\bar{x}_1 + i \bar{x}_2)/\sqrt{2}$. By using this and the relation for bosons $\exp(\mu \hat{a})\exp(\nu \hat{a}^{\dagger} \hat{b}^{\dagger}) = \exp(\mu\nu \hat{b}^{\dagger}) \exp(\nu \hat{a}^{\dagger}\hat{b}^{\dagger}) \exp(\mu \hat{a})$ derived from the Baker-Hausdorff formula \cite{Scully and Zubairy}, we find
\vspace{-2mm}
\begin{eqnarray}
\langle \bar{x}_1, \phi|_{s1} \langle \bar{x}_2, \phi+\mbox{$\frac{\pi}{2}$}|_{s2} | \eta e^{i2\phi} \rangle_{1,2} = e^{-\frac{|\gamma|^2}{2}}\sqrt{\frac{1-\eta^2}{2\pi}} \hspace{1cm} \nonumber\\
\times \exp(-\eta \gamma e^{i\phi} \hat{a}_2^\dagger ) \left( \sum_{n=0}^{\infty} \eta^n  |n \rangle_2 \langle n |_{in} \right) \exp(\gamma^* e^{-i\phi} \hat{a}_{in}). \quad \label{CVQT: quadrature eigenstates}
\end{eqnarray}

With orthogonality approximation of the coherent state, we find $\hat{\rho}_\mathrm{II}$ includes Eq.\ (\ref{CVQT: quadrature eigenstates}).
Ideal quantum teleportation is possible only when $\eta = 1$, where a two-mode squeezed state is maximally entangled \cite{CVQT}.
Eq.\ (\ref{CVQT: quadrature eigenstates}) shows that the unitary transform $\hat{U}_2$ applied by Bob to the mode $2$ in the reconstruction step must satisfy
$
 \hat{U}_2 |_{\eta=1} \exp(\mbox{\scriptsize $-\frac{|\gamma|^2}{2}$}) \exp(-\gamma e^{i\phi} \hat{a}_2^\dagger ) \exp(\gamma^* e^{-i\phi} \hat{a}_2)= const.
$,
because $\sum_{n=0}^{\infty} |n \rangle_2 \langle n |_{in}$ transfers a state of the mode $in$ to the mode $2$ with absolute precision.

The necessary condition for $\hat{U}_2$ is then found to be $\hat{U}_2 |_{\eta=1} = \exp ( \gamma e^{i\phi} \hat{a}_2^\dagger - \gamma^* e^{-i\phi} \hat{a}_2 )$, which means Bob needs not only the measurement results by Alice $\gamma$ but also the unknown phase of Alice's LO fields $\phi$ to perform $\hat{U}_2$. Hence, to share $\phi$ between Alice and Bob by a certain means is essential to realizing CVQT with a laser.

In the experiment \cite{CVQT Experiment}, Bob obtains $\phi$ from the LO at hand directly connected to Alice's LOs and the pump field of a two-mode squeezed state. If Bob performs the unitary transform $\hat{U}_{2,l3}(\gamma, \eta) \equiv \exp [ (\eta/r_o) (\gamma \hat{a}_{l3} \hat{a}_2^\dagger
- \gamma^* \hat{a}_{l3}^\dagger \hat{a}_2 ) ]$ after he obtains $\gamma$, the density operator of the total system $\hat{\rho}_\mathrm{III} \equiv \hat{U}_{2,l3}(\gamma, \eta) \hat{\rho}_\mathrm{II} \hat{U}_{2, l3}^\dagger (\gamma, \eta)$ becomes
\vspace{-2mm}
\begin{eqnarray}
\lefteqn{\hat{\rho}_\mathrm{III} \sim P^{-1}(\bar{x}_1, \bar{x}_2)} \nonumber\\
&\times& \!\! \int_0^{2\pi} \!\! \frac{d\phi}{2\pi} \  |r_o e^{i\phi} \rangle_{l1} |\bar{x}_1, \phi \rangle_{s1} |r_o e^{i(\phi+\frac{\pi}{2})} \rangle_{l2} |\bar{x}_2, \phi+\mbox{$\frac{\pi}{2}$} \rangle_{s2}  \nonumber \\
& & \otimes |r_o e^{i\phi} \rangle_{l3} \hat{T}_{2,in}(\gamma, \eta, \phi) \, \hat{\rho}_{in} \, \hat{T}_{2,in}^\dagger (\gamma, \eta, \phi) \langle r_o e^{i\phi} |_{l3} \nonumber\\
& & \otimes \langle \bar{x}_2, \phi+\mbox{$\frac{\pi}{2}$} |_{s2} \langle r_o e^{i(\phi+\frac{\pi}{2})} |_{l2} \langle \bar{x}_1, \phi |_{s1} \langle r_o e^{i\phi} |_{l1}, \quad \label{CVQT: final density operator}
\end{eqnarray}
where $\hat{a}^\dagger|r_o e^{i\theta} \rangle \sim r_o e^{-i\theta} |r_o e^{i\theta} \rangle$ in the limit $r_o\to+\infty$ and $\hat{T}_{2,in}$ is defined as
\vspace{-2mm}
\begin{eqnarray}
\hat{T}_{2,in} (\gamma, \eta, \phi)
\equiv e^{-\frac{|\gamma|^2}{2}(1-\eta^2)} \sqrt{\frac{1-\eta^2}{2\pi}} \hspace{30mm}  \nonumber\\
\times \exp(-\eta \gamma^* e^{-i\phi} \hat{a}_2 ) \! \left( \sum_{n=0}^{\infty} \eta^n |n \rangle_2 \langle n |_{in} \right) \! \exp(\gamma^* e^{-i\phi} \hat{a}_{in} ), \quad  \label{CVQT: transfer operator}
\end{eqnarray}
which corresponds to the transfer operator in Ref.\ \cite{Hofmann} from the mode $in$ to the mode $2$.

Eqs.\ (\ref{CVQT: final density operator}) and (\ref{CVQT: transfer operator}) clearly show that in the special case $\eta=1$ $\hat{T}_{2,in}$ is independent of the unknown phase $\phi$ where ideal quantum teleportation is realized, while in the usual case $0 \leq \eta < 1$ $\hat{T}_{2,in}$ is dependent on the unknown phase $\phi$ where the reconstructed density operator in the mode $2$ is distorted from $\hat{\rho}_{in}$.

We will subsequently discuss generation of a strongly phase-correlated quantum state necessary in CVQT by measuring two independent laser fields
\begin{eqnarray}
\hat{\rho}_o \! = \!\! \int_0^{2\pi} \!\!\! \frac{d\phi_a}{2\pi} \!\! \int_0^{2\pi} \!\!\! \frac{d\phi_b}{2\pi}
 |r_a e^{i\phi_a} \rangle_a |r_b e^{i\phi_b} \rangle_b
\langle r_b e^{i\phi_b}|_b \langle r_a e^{i\phi_a} |_a.  \quad \label{BHD for lasers: initial density operator}
\end{eqnarray}

In the case of BHD, since the observable satisfies $(\hat{a}^\dagger\hat{b}+\hat{a}\hat{b}^\dagger)|r_a e^{i\phi}\rangle |r_b e^{i(\phi \pm \varphi)}\rangle  \!\sim\!  2r_a r_b \cos(\varphi) |r_a e^{i\phi}\rangle |r_b e^{i(\phi \pm \varphi)}\rangle \; (0 \le \varphi \le \pi)$ in the limit $r_a, r_b \!\to\! +\infty$, the measurement operator for OBPM may be defined as
$
\hat{M}(\overline{\cos(\varphi)}, r_1, r_2, \phi)
\equiv \pi^{-1} |r_1 e^{i\phi}\rangle |r_2 e^{i(\phi \pm \varphi)}\rangle
\langle r_2 e^{i(\phi \pm \varphi)}| \langle r_1 e^{i\phi}|
$.
Then, the density operator after the measurement becomes
$
\hat{\rho} = \frac{1}{2} \> \int_0^{2\pi} \! \frac{d\phi}{2\pi} \; |r_a e^{i\phi} \rangle_a |r_b e^{i(\phi+\varphi)} \rangle_b \langle r_b e^{i(\phi+\varphi)}|_b \langle r_a e^{i\phi} |_a + \frac{1}{2} \> \int_0^{2\pi} \! \frac{d\phi}{2\pi} \; |r_a e^{i\phi} \rangle_a |r_b e^{i(\phi-\varphi)} \rangle_b \langle r_b e^{i(\phi-\varphi)}|_b \langle r_a e^{i\phi} |_a$,
i.e., BHD does not determine a unique phase difference of two lasers except exactly when $\overline{\cos(\varphi)}=\pm1$ with negligible probability. Hence, the generated quantum state by BHD is not applicable to CVQT to share the unknown phase of the laser field.

\begin{figure}
\includegraphics[width=6cm]{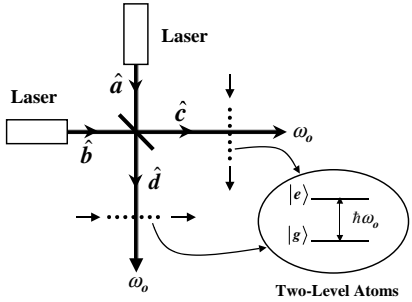}
\caption{\label{fig:setup} Experimental setup for continuous measurement of two independent laser fields. $\hat{a}, \hat{b}$ and $\hat{c}, \hat{d}$ are annihilation operators for the input and output modes of a 50/50 beamsplitter, respectively, satisfying $\hat{c}=(\hat{a}-\hat{b})/\sqrt{2},\; \hat{d}=(\hat{a}+\hat{b})/\sqrt{2}$. Two-level atoms resonant with the laser fields are all prepared in the ground state beforehand, and go across the output fields one by one at regular intervals.}
\end{figure}

But if we perform continuous measurement \cite{Carmichael, DCM, Scully and Zubairy} presented in Fig.\ \ref{fig:setup}, a unique phase difference of two lasers is chosen with non-zero probability. In Fig.\ \ref{fig:setup}, two-level atom beams are used as probes to ensure that photoabsorption occurs at most one time within the infinitesimal atom-field interaction time $\tau$, which is not feasible by a present photodetector lacking single photon resolution in the strong field \cite{Carmichael, Gardiner and Zoller}.

Given that the total photoabsorption (quantum jump) occurs either in the mode $c$ or $d$ at times $t_1, t_2, \ldots, t_s$ in the time interval $[0, t]$ with no absorption between these times, the conditional probability that photoabsorption occurs $p, q(=\!s\!-\!p)$ times in the mode $c, d$, respectively, is
\vspace{-3.5mm}
\begin{eqnarray}
\hspace*{-4.7mm}P(t; p,q|s) \!\! &\simeq& \!\! \frac{ {s \choose p} \mathrm{Tr}\{\hat{c}^p \hat{d}^q e^{-R (\hat{c}^\dagger \hat{c} + \hat{d}^\dagger \hat{d})t} \hat{\rho}(0) e^{-R (\hat{c}^\dagger \hat{c} + \hat{d}^\dagger \hat{d})t} \hat{d}^{\dagger q} \hat{c}^{\dagger p} \} }
{\!\!{\displaystyle \!\! \sum_{p+q=s}} \!\!\! {s \choose p} \mathrm{Tr}\{\hat{c}^p \hat{d}^q e^{-R (\hat{c}^\dagger \hat{c} + \hat{d}^\dagger \hat{d})t} \hat{\rho}(0) e^{-R (\hat{c}^\dagger \hat{c} + \hat{d}^\dagger \hat{d})t} \hat{d}^{\dagger q} \hat{c}^{\dagger p} \} } \nonumber\\
&=& \pi^{-1} {s \choose p} B(p+\mbox{$\frac{1}{2}$}, q+\mbox{$\frac{1}{2}$}),  \label{CM: conditional probability for absorption}
\end{eqnarray}
and the density operator becomes
\vspace{-1mm}
\begin{eqnarray}
\hspace*{-3.5mm}\hat{\rho}(t; p, q) \!\! &\simeq& \!\! \frac{\hat{c}^p \hat{d}^q e^{-R (\hat{c}^\dagger \hat{c} + \hat{d}^\dagger \hat{d})t} \hat{\rho}(0) e^{-R (\hat{c}^\dagger \hat{c} + \hat{d}^\dagger \hat{d})t} \hat{d}^{\dagger q} \hat{c}^{\dagger p}} {\mathrm{Tr}\{ \hat{c}^p \hat{d}^q e^{-R (\hat{c}^\dagger \hat{c} + \hat{d}^\dagger \hat{d})t} \hat{\rho}(0) e^{-R (\hat{c}^\dagger \hat{c} + \hat{d}^\dagger \hat{d})t} \hat{d}^{\dagger q} \hat{c}^{\dagger p} \}} \hspace{9mm} \nonumber\\
&=& \!\! \frac{\pi}{B(p+\frac{1}{2}, q+\frac{1}{2})} \! \int_0^{2\pi} \!\!\! \frac{d\phi_a}{2\pi} \!\! \int_0^{2\pi} \!\!\! \frac{d\phi_b}{2\pi} \ \sin^{2p} \bigl( \mbox{$\frac{\phi_a-\phi_b}{2}$} \bigr) \nonumber\\
&\times& \!\! \cos^{2q} \bigl( \mbox{$\frac{\phi_a-\phi_b}{2}$} \bigr) |r_t e^{i\phi_a} \rangle_a |r_t e^{i\phi_b} \rangle_b \langle r_t e^{i\phi_b}|_b \langle r_t e^{i\phi_a} |_a , \quad \label{CM: density operator}
\end{eqnarray}
where $\rho(0)$ is Eq.\ (\ref{BHD for lasers: initial density operator}) with $r_a\!=\!r_b \> (\equiv \! r_o)$, $e^{-R\hat{d}^\dagger\hat{d}(p\tau)} \sim e^{-R\hat{c}^\dagger\hat{c}(q\tau)} \sim 1$, $B(x, y)$ is the beta function, $r_t \equiv r_o e^{-Rt}$, $R \equiv g^2 \tau /2$, and $g$ is the atom-field coupling constant \cite{Scully and Zubairy}.

The proposed continuous measurement is valid when $\tau \ll (\sqrt{2s}\,r_o g)^{-1}$.
In Eq.\ (\ref{CM: density operator}), we find that atoms simultaneously intersecting the output modes with no absorption (null measurement) damp both laser fields, leaving phase correlation between the fields unchanged.
The absorption rate is assumed to be quite high, where $t$ is much smaller than the dynamical time scale of an individual laser.

For $s \gg 1$, the distribution of the phase difference of states in the integrand of Eq.\ (\ref{CM: density operator}) has a peak at $|\phi_a-\phi_b|=\pi$ when $p=s$, or at $|\phi_a-\phi_b|=0$ when $p=0$. Since Eq.\ (\ref{CM: conditional probability for absorption}) has peaks at $p=0, s$, the probability of obtaining Eq.\ (\ref{CM: density operator}) with $p=0, s$ is not negligible.

\begin{figure}
\includegraphics[width=8cm]{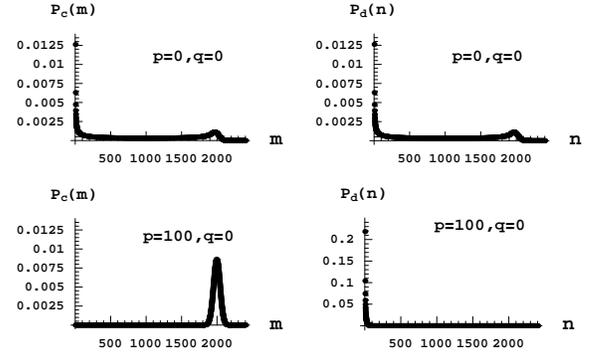}
\caption{\label{fig:photon_distribution} Photon number distributions from Eq.\ (\ref{CM: photon number distribution}) for $p=s$ with ${r_t}^2=10^3$. Given that $s=100$, the probability Eq.\ (\ref{CM: conditional probability for absorption}) for $p=0$ or $p=100$ is about $11.3\%$. If the Monte Carlo wave-function procedure \cite{DCM} is performed, gradual decay of $r_t$ due to null measurement shall be seen besides the above distribution change. $P_c(m)$ approaches a Poisson distribution as $s$ becomes large.}
\end{figure}

The photon number distribution of the mode $c$, $P_c(m) \equiv \langle m| \mathrm{Tr}_d\{\hat{\rho}(t; p, q)\} |m \rangle_c$, is found to be\vspace{-1mm}
\begin{eqnarray}
P_c(m) &=& e^{-2{r_t}^2} \frac{\bigl( 2 {r_t}^2 \bigr)^m}{m!} \frac{B(m+p+\frac{1}{2},q+\frac{1}{2})}{B(p+\frac{1}{2},q+\frac{1}{2})}\nonumber \\
& & \hspace{5mm} \times {}_1 F_1(q+\mbox{$\frac{1}{2}$}; m+p+q+1; 2 {r_t}^2) ,\quad \label{CM: photon number distribution}
\end{eqnarray}
where ${}_1F_1(\alpha; \beta; z)$ is the confluent hypergeometric function of the first kind. $P_d(n)$ is easily obtained by replacing $m$ with $n$ and interchanging $p \leftrightarrow q$ in Eq.\ (\ref{CM: photon number distribution}).
Fig.\ \ref{fig:photon_distribution} is for $P_c(m), P_d(n)$.

When $p=0,s$ with $s \gg 1$, the generated quantum state is applicable to CVQT as a means to share the unknown phase of the laser field between Alice and Bob, though the phase correlation formed after the continuous measurement will slowly be broken by the phase diffusion effect of lasers.

The famous experiment for interference of two independent lasers by Pfleegor and Mandel \cite{PM}, where {\it weak laser fields} were mixed by beamsplitters and all the output fields were continuously measured by photomultipliers, should carefully be reviewed in terms of phase-correlated quantum state generation by measurement.

In conclusion, we have shown that, contrary to the claims of Ref.\ \cite{vEF}, the standard description of the laser field used in Ref.\ \cite{RS} is sufficient to understand CVQT with a conventional laser.
By using the standard description of the laser field \cite{RS}, we have presented OBPM for BHD to analyze CVQT with a laser. CVQT is found to be possible only if the unknown phase of the laser field is shared among Alice's LOs, the EPR state, and Bob's LO by a certain means. The demonstrated experiment for CVQT \cite{CVQT Experiment} is valid, but needs an optical path other than the EPR channel and a classical channel allowed to use in the teleportation protocols \cite{Original QT, CVQT} to share the unknown phase of the same laser field between Alice and Bob. We have proposed a method to probabilistically generate a strongly phase-correlated quantum state via continuous measurement of independent lasers, which is applicable to realizing CVQT without the additional optical path.

The author is greatly indebted to Kenji Ohkuma, Kouichi Ichimura, and Noritsugu Shiokawa for intensive and fruitful discussions on the subject of this Letter. The author also acknowledges useful discussions with Mio Murao, Hirofumi Muratani, and Toshiaki Iitaka.

\end{document}